\documentclass[a4paper,aps,pra,reprint,twocolumn,superscriptaddress]{revtex4-1} 

\usepackage[utf8]{inputenc} 
\usepackage{textcomp} 
\usepackage{graphicx}  
\usepackage{amsmath,amssymb}  
\usepackage{bm}  
\usepackage{rotating}

\begin{document}

\title{Direct generation of optical frequency combs 
in $\chi^{(2)}$ nonlinear cavities
}

\author{S.~Mosca}
\author{I.~Ricciardi}
\author{M.~Parisi}
\author{P.~Maddaloni}
\author{L.~Santamaria}
\affiliation{CNR-INO, Istituto Nazionale di Ottica, Via Campi Flegrei 34, 80078 Pozzuoli (NA), Italy}
\author{P.~De~Natale}
\affiliation{CNR-INO, Istituto Nazionale di Ottica, largo E. Fermi 6, 50125 Firenze, Italy}
\author{M.~De~Rosa}
\email[Corresponding author: ]{maurizio.derosa@ino.it}
\affiliation{CNR-INO, Istituto Nazionale di Ottica, Via Campi Flegrei 34, 80078 Pozzuoli (NA), Italy}

\begin{abstract}
{Quadratic nonlinear processes are currently exploited  for frequency comb transfer and extension from the visible and near infrared regions to other spectral ranges where direct comb generation cannot be accomplished. 
However, frequency comb generation has been directly observed in continuously-pumped quadratic nonlinear crystals placed inside an optical cavity. 
At the same time, an introductory theoretical description of the phenomenon has been provided, showing a remarkable analogy with the dynamics of third-order Kerr microresonators.  
Here, we give an overview of our recent work on $\chi^{(2)}$ frequency comb generation. Furthermore, we generalize the preliminary three-wave spectral model to a many-mode comb and present a stability analysis of different cavity field regimes. 
Although at a very early stage, our work lays the groundwork for a novel class of highly efficient and versatile frequency comb synthesizers based on second-order nonlinear materials.}

\end{abstract}

\maketitle

\section{Introduction}

The advent of optical frequency combs (OFCs), consisting of thousands of equally spaced sharp laser frequencies, has totally revolutionized the field of absolute frequency metrology, providing a direct link between the optical and radio frequency (RF) regions of the electromagnetic spectrum~\cite{Udem:2002bj}.
The huge and complicated frequency chains used to bridge the optical-to-RF gap became obsolete, making extremely precise optical frequency measurements accessible to many laboratories.
Originally conceived for frequency metrology, nowadays frequency combs represent a fast-growing research field, as they are routinely used in a wide range of scientific and technological applications, from synchronization of telecommunication systems to astronomical spectral calibrationm from biomedical to environmental spectrometry.

OFCs naturally arise as the spectral counterpart of a regular pulsed laser emission~\cite{Siegman:Lasers}. 
Indeed, mode-locked lasers have been originally investigated as a suitable tool for comb generation, culminating, at the turn of the millennium, in the octave spanning emission needed for absolute frequency referencing~\cite{Diddams:2000kv,Jones:2000tn,Holzwarth:2000aa}, also recognized with the 2005 Nobel Prize in Physics~\cite{Hall:2006by,Hansch:2006el}. 
In 2007, frequency combs have been demonstrated in continuously pumped nonlinear passive microresonantors, exploiting third-order, $\chi^{(3)}$, parametric processes~\cite{DelHaye:2007gi}.  
In such Kerr microresonators, cascaded four-wave mixing (FWM) processes lead to a uniform broadband frequency comb, whereas self- and cross-phase modulation (SPM and XPM) compensate the unequal spacing of the cavity modes due to the group velocity dispersion (GVD) of the material.
More recently, mid-infrared OFCs have been observed in quantum cascade lasers, both in the mid-infrared and terahertz ranges~\cite{Hugi:2012ep,Burghoff:2014hp,Lu:2015en}, and attempts of modeling their dynamics are in progress~\cite{Friedli:2013id,Khurgin:2014hy,Villares:2015ho,Cappelli:2015aa}.

Materials with second-order susceptibility, $\chi^{(2)}$, have been used to indirectly replicate or extend an existing frequency comb, otherwise generated, to different spectral regions, exploiting different $\chi^{(2)}$ processes, like  difference frequency generation~\cite{Maddaloni:2006ka,Maddaloni:2009bg,Galli:2013cg,Galli:2014dt,Erny:2007dj,Gambetta:2008fa,Gambetta:2013ci,Keilmann:2012bb}, parametric generation in synchronously pumped optical parametric oscillators (OPOs)\cite{Sun:2007wu,Wong:2008fz,Adler:2009ka,Wong:2010gw,Leindecker:2011ch}, and harmonic up-conversion~\cite{Peters:2009ab,Kandula:2010ab,Bernhardt:2012ab}. 
Moreover, frequency comb generation has also been directly observed in continuously-pumped  quadratic nonlinear crystals placed inside an optical cavity. 
A first attempt to observe direct frequency comb generation in $\chi^{(2)}$ media was made by Diddams and co-workers, using a nearly degenerate, continuously pumped OPO with intracavity phase modulation similar to actively mode-locked lasers~\cite{Diddams:1999ch}. 
Ulvila et al. used a singly resonant OPO with an additional intracavity crystal, observing a comb around the signal wavelength, which was then replicated around the idler wavelength by difference frequency with the pump~\cite{Ulvila:2013jv,Ulvila:2014bx}.
We observed frequency comb generation in a single-crystal nonlinear cavity: frequency combs appear both around the fundamental pump frequency and its second harmonic for a wide range of phase matching conditions. We explained the comb generation as the result of a series of cascaded $\chi^{(2)}$ processes, presenting a simple but insightful theoretical model~\cite{Ricciardi:2015bw}.  

In this Article, we review our work on direct generation of OFCs in quadratic materials. 
Although at a very early stage, $\chi^{(2)}$ frequency combs are deeply rooted in a long-standing study about cascaded quadratic processes. Nevertheless, they represent a new promising paradigm of comb generation, paving the way to a novel class of highly efficient and versatile frequency comb synthesizers, possibly in miniaturized devices~\cite{Ilchenko:2004eo,Furst:2010bt,Furst:2010co,Levy:2011wp,Cazzanelli:2012jz,Jung:2014jt,Miller:2014dz,Kuo:2014aa,Mariani:2014gw}. 
Combs in Kerr microresonators or quantum cascade lasers are suitable candidates for miniaturized integrated devices with potential application in fields where size, power consumption and low cost are important.
We show that $\chi^{(2)}$ combs, inherently more efficient than Kerr combs, when configured in micro- or nanometric structures can be operated with pump powers from milliwatt to microwatt scale, taking advantage of  different materials with a strong quadratic nonlinearity over a wide spectral range~\cite{Levy:2011wp,Cazzanelli:2012jz,Jung:2014jt,Miller:2014dz,Kuo:2014aa,Mariani:2014gw}. 

In Section \ref{sec:cascade}, we shortly recall the basic description of second-order nonlinear three-wave interaction and discuss the fundamental aspects of cascaded nonlinear processes.
Afterward, we give an overview of the techniques adopted to transfer OFCs, typically generated in the visible and near infrared (NIR) range, to MIR, UV and Terahertz spectral regions (Sec.~\ref{sec:transfer}). Then, in Sec.~\ref{sec:chi2FC}, we review the fundamental results about OFC generation in $\chi^{(2)}$ systems with cw-pumping.
Thereafter, we extend the analysis of the $\chi^{(2)}$ system presented in~\cite{Ricciardi:2015bw}, generalizing the effective $\chi^{(3)}$ picture to a generic number of interacting sub-harmonic fields (Sec.~\ref{sec:modal-eq}). 
We finally investigate the stability regimes of the singly resonant nonlinear cavity: below threshold for cascaded OPO, revisiting the bistable  regimes previously observed in singly resonant cavity SHG~\cite{White:1996ts}; and above threshold, deriving an approximate analytical expression for the comb power threshold.

\section{Cascaded $\chi^{(2)}\!\!:\!\chi^{(2)}$ processes}
\label{sec:cascade}
An electromagnetic field passing through a dielectric material induces oscillations of the atomic dipoles which contribute to the  polarization $P(t)$ of the material, i.e., the dipole moment per unit volume. 
In linear approximation, valid for weak field amplitude, the dipoles oscillate at the same frequency $\omega$ of the applied field and the polarization is proportional to the field amplitude.
As the field intensity increases the inherent anharmonic response of the atomic dipoles, related to the motion of bound electrons, cannot be neglected: terms higher than first-order become relevant and, as a result,  the induced polarization can be expressed as powers of the electric field amplitude, which in a scalar form reads
\[
P(t) =\epsilon_0 \, [ \chi^{(1)} E(t) +  \chi^{(2)} E^2 (t) +\chi^{(3)} E^3 (t)+ ... ]  \, ,
\]
where $\epsilon_0$ is the vacuum permittivity and $\chi^{(j)}$ is the $j$-th order susceptibility. More generally, considering the vector nature of the fields, $\chi^{(j)}$ is a $(j+1)$-th rank tensor.

Quadratic nonlinearity, governed by the second-order susceptibility $\chi^{(2)}$, leads to three-wave interactions where, in a simplified picture, two input optical fields at frequencies $\omega_1$ and $\omega_2$ incident upon the nonlinear medium give rise to nonlinear polarization components, at new frequencies: second harmonics, $2\omega_1$ and $2\omega_2$; sum-frequency, $\omega_1+\omega_2$; and difference frequency, $\omega_1-\omega_2$, or optical rectification~\cite{Boyd:NLO}.

The achievement of high conversion efficiency is a primary goal of most nonlinear devices; however, no more that one component typically results in a noticeable intensity at the end of the medium, unless a particular phase relation between the interacting waves is satisfied, allowing the radiation emitted by dipoles from different regions along the path to interfere constructively and add up efficiently.
In fact, because of chromatic dispersion of the material, the interacting waves actually propagate with different phase velocities and light generated in different regions over twice the coherence length $l_c= \pi/\Delta k$ will be out of phase and interfere destructively (here, $\Delta k = k_3-k_2-k_1$ is the mismatch wave vector and depends on the wave vectors of the interacting fields). As a result, in every three-wave process, the energy flow repetitively reverses the direction, going back and forth amongst the interacting fields as the beams propagate through the medium. 
When a phase matching condition is satisfied, i.e., $\Delta k =0$, constructive interference between nonlinear polarization and radiated field occurs all along the interaction length, leading to an efficient unidirectional conversion into the output field.
A phase matching condition can be achieved in birefringent crystals by properly choosing the polarization and the propagation direction of the interacting waves with respect to the crystal optical axes. Otherwise, the nonlinear susceptibility of the medium can be artificially engineered by spatial inversion of ferroelectric domains in the so called periodically poled crystals. 
While simultaneous phase matching of several  $\chi^{(2)}$ processes cannot be achieved in a single birefringent crystals, except for a fortuitous coincidence~\cite{Pfister:2004fs}, domain engineering, conversely, makes possible to realize simultaneous phase matching conditions~\cite{Pooser:2005cy}.

Nevertheless, locally, multiple $\chi^{(2)}$ processes actually take place and waves of different frequencies are generated which, however small, can in turn be the input for new cascaded $\chi^{(2)}\!\!\!:\!\!\chi^{(2)}$ processes as they propagate through the medium.  
Importantly, cascaded $\chi^{(2)}\!\!\!:\!\!\chi^{(2)}$ processes can mimic third-order processes as predicted as early as in the late sixties of the 20th century.
The observation of self-action effects, commonly identified with cubic nonlinearity, was predicted for quadratic nonlinear crystals by L. A. Ostrovskii, who envisaged amplitude dependent refractive index and self-focusing in an off-phase matched frequency doubling crystal~\cite{Ostrovskii:1967}. 
It took more than two decades to set up an experimental measurement of the nonlinear phase shift, which stimulated, in turn, an increasing interest towards cascading effects~\cite{Desalvo:1992cs}.
The nonlinear refractive effect in an off-phase matched SH medium is due to a portion of the frequency-doubled light which is down-converted back to the fundamental frequency with a shifted phase. As a result, the net phase of the fundamental wave is shifted in proportion to the irradiance of the fundamental field, resulting in a Kerr-like nonlinearity. 
It is worth noting that phase mismatch is not an essential requirement for Kerr-like nonlinearity~\cite{Saltiel:2005te}, as, for instance, in the cascaded sequence of phase-matched SHG and SFG between the SH and the fundamental field, finally producing the third harmonic of the fundamental~\cite{Ricciardi:2009us}.

\begin{figure}[t!]
\begin{center}
\includegraphics*[bbllx=0bp,bblly=0bp,bburx=440bp,bbury=365bp,width=75mm]{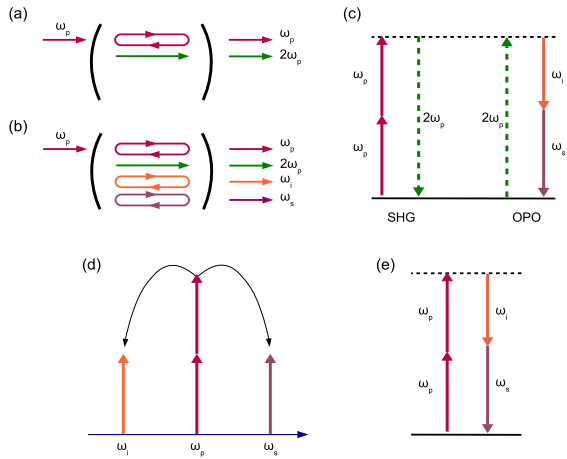}
\caption{Internally pumped OPO in intracavity SHG. (a) fundamental and second harmonic fields in a singly resonant SHG cavity; (b) for sufficient SH power an internally pumped nondegenerate OPO occurs, clamping the SH power to a constant level; (c) energy-level diagram for the sequence of second harmonic generation and optical parametric oscillation: in this regime there is no net SH generation; (d) in the subharmonic range, around the fundamental frequency, the process can be seen as a degenerate FWM, where two pump photons are converted in two frequency symmetric parametric photons; (e) energy-level diagram for effective degenerate FWM.}
\label{fig:SHG+OPO}
\end{center}
\end{figure}

A cascaded process of primary interest for the following discussion is the internally pumped (or subharmonic-pumped) optical parametric oscillator, occurring in cavity-enhanced SHG, where a SH phase-matched crystal is placed in an optical cavity, resonant for a frequency range around the fundamental pump frequency $\omega_1$.
As the pump power is increased, the harmonic power generated within the nonlinear crystal can exceed the threshold power for SH-pumped parametric oscillation, leading to a cascaded OPO, with steady oscillations of a frequency-symmetric signal-idler (s-i) pair around the fundamental frequency (Fig.~\ref{fig:SHG+OPO}). 
As the laser power exceeds the threshold value, the second-harmonic power clamps at a constant value, irrespective of the increasing input power, implying a net photon (or energy) transfer from pump to parametric field, mediated by ``virtual'' SH photons. 
Aside from the SH field the cascaded SHG+OPO is analogue to degenerate four-wave mixing in $\chi^{(3)}$ media. 
We notice that, although the SHG phase-matched nonlinear cavity is ideally phase matched also for the inverse process of degenerate OPO, this particular process cannot take place due to the relative phase condition imposed by the primary SHG, $\phi_2 - 2\phi_1 =-\pi/2$, while degenerate OPO requires $\phi_2 - 2\phi_1 =\pi/2$~\cite{Armstrong:1962wi}. On the other hand, nondegenerate parametric oscillation with two new distinct parametric fields is allowed as their phases are free to adjust with respect to the driving SH  field phase.

Subharmonic pumped OPO was first observed in a monolithic cavity resonating for both the fundamental and harmonic waves~\cite{Schiller:1993tz}; subsequently, it was implemented in a simpler cavity, resonant only for the fundamental frequency~\cite{Schiller:1996gx,Schneider:1997ks,White:1997ta}. 
It was possible to observe simultaneous parametric oscillation of two s-i pairs around the fundamental frequency, while, in the harmonic region, the evidence of sum frequency processes between the fundamental and each parametric  wave was observed, showing the wealth of phenomena which are triggered by the first cascaded three-wave mixing.

A theoretical model was derived, based on the reduced set of coupled mode equations for the two $\chi^{(2)}$ processes of frequency doubling and cascaded degenerate OPO~\cite{Schiller:1997dp}.
A perturbative solution gives, for the three resonating sub-harmonic fields, a set of dynamic equations which displays effective third-order interaction terms. However, not all the relevant terms appear in these equations. 
A more complete expression can be found by considering a larger set of coupled mode equations, including generation of the signal (idler) second harmonic and sum frequency of signal (idler) and fundamental. Once these processes are considered, the derived dynamic equations are still limited to three sub-harmonic fields, but with new relevant interaction terms~\cite{Ricciardi:2015bw}. 

Phase mismatched cavity enhanced SHG, below the threshold for the onset of the cascaded OPO, displayed optical bistability~\cite{White:1996ts} and phase shift~\cite{Ou:1995kx} for the fundamental wave, typically of cubic nonlinearity. In this case, the systems are nothing else than cavity-embedded versions of single pass phase mismatched SH crystals~\cite{Desalvo:1992cs}.
The analogy of a  phase-mismatched cavity SHG with a Kerr cavity drew the attention on the quantum properties of such a system as a way to generate squeezed states of light ~\cite{White:1997ta,White:2000ir}. 
More recently, a similar nonlinear cavity was used as an effective Kerr cavity for power noise reduction of a cw laser~\cite{Khalaidovski:2009ga}.

\section{OFC transfer/extension with $\chi^{(2)}$ systems}
\label{sec:transfer}

Frequency combs are directly available in the wavelength region from 400~nm to 2.2~$\mu$m and have played a crucial role for the development of new ultraprecise optical atomic clocks and precision spectroscopy. Besides visible and near-infrared spectral domains, the mid-infrared region is very attractive for the presence of strong characteristic vibrational transitions of a large number of molecules. Thus, frequency combs have been transferred to the mid-infrared region through different nonlinear frequency conversions in $\chi^{(2)}$ media~\cite{Schliesser:2012dn,Adler:2010da}.
MIR frequency combs from 2.9 to 3.5~$\mu$m~\cite{Maddaloni:2006ka} and in the 4.2-5.0~$\mu$m range~\cite{Galli:2013cg} have been realized by difference frequency generation (DFG) between a near-infrared femtosecond comb and a continuous-wave source.   
Also, by DFG between distinct teeth of the same comb, NIR frequency combs have been transferred in the 3.2-4.8~$\mu$m range~\cite{Erny:2007dj} and at longer MIR wavelengths up to 14~$\mu$m~\cite{Gambetta:2008fa,Gambetta:2013ci}.
Finally, a MIR source from 4 up to 17~$\mu$m was realized by using DFG between supercontinuum pulses of the same comb~\cite{Keilmann:2012bb}.
An alternative and more efficient technique for generating phase- and frequency-locked frequency combs in the MIR is based on optical parametric oscillators (OPOs), which both down-convert and augment the spectrum of a pump comb in the near-IR~\cite{Leindecker:2011ch,Adler:2009ka}. Femtosecond frequency combs have been realized around 1550~nm by use of self-phase-locked degenerate synchronously pumped OPO~\cite{Wong:2008fz,Wong:2010gw}, while a comb from 0.4 to 2.4~$\mu$m was obtained by phase locking a Ti:sapphire laser and a synchronously pumped OPO to a common supercontinuum reference~\cite{Sun:2007wu}.

In addition to MIR, femtosecond frequency combs techniques have been extended up to the extreme ultraviolet region, highly interesting for novel precision QED tests and direct two-photon atomic spectroscopy. A frequency comb at 205 nm has been generated by frequency quadrupling a Ti:sapphire laser at 820 nm~\cite{Peters:2009ab}, while, by amplification and coherent upconversion of a pair of near-infrared pulses, the femtosecond comb has been transferred to XUV wavelengths near 51~nm~\cite{Kandula:2010ab}. An ultraviolet comb up to 57~nm has been realized, too, by a cavity-assisted frequency-doubling of an Yb fiber laser and high harmonics generation in a xenon gas jet~\cite{Bernhardt:2012ab}.

Besides bulk crystals, periodically poled lithium niobate waveguides have been used for octave spectral broadening of fs Er-, Yb- and Tm-doped fiber lasers~\cite{Langrock:2007eg,Phillips:2011hv}. Furthermore, by optical rectification of a fs Ti:sapphire laser in a MgO-doped LiNbO$_3$ waveguide, the metrological features of optical frequency comb synthesizers have been extended to the terahertz domain~\cite{Consolino:2012eq,Bartalini:2014jb}. 
Interestingly, in addition to $\chi^{(3)}$ nonlinearity, Kerr microresonators can present second order nonlinearity, whether intrinsic to the material, as in AlN waveguides~\cite{Jung:2013ju}, or induced by proper material engineering of the original centrosymmetric structures~\cite{Cazzanelli:2012jz,Levy:2011wp}. In this kind of microresonators $\chi^{(2)}$ nonlinearity was exploited to frequency up-convert the original $\chi^{(3)}$ comb in the second- and third harmonic ranges~\cite{Jung:2014jt,Miller:2014dz}.

\section{OFC generation in cw-pumped $\chi^{(2)}$ system}
\label{sec:chi2FC}

In 1999 Diddams and coworkers proposed a different approach to traditional mode-locking of lasers for obtaining optical frequency combs~\cite{Diddams:1999ch}, implementing an OFC generator based on active phase-modulation in a continuously pumped OPO cavity. The OPO, pumped by single frequency light at 532 nm, produced near-degenerate signal and idler fields; when an intracavity electro-optic modulator was driven at a frequency equal to the cavity FSR, a comb structure about both the signal and the idler was generated, resulting in a dispersion-limited comb with 20~nm spanning.

The possibility of frequency comb generation in a cw-pumped $\chi^{(2)}$ system without any active modulation was first reported by Ulvila et al.~\cite{Ulvila:2013jv}. They used a singly resonant OPO with two periodically-poled LiNbO$_3$ crystals: the first crystal was designed for parametric oscillation of signal and idler around 1.56~$\mu$m and 3.35~$\mu$m, respectively, when pumped by a 1064~nm pump laser; the second crystal poling period provided phase mismatched SHG for the resonating signal. Above the OPO threshold, spectral analysis of the signal beam leaking from the cavity showed a broadened spectral emission around the signal frequency. RF beat note at the FSR of the OPO cavity confirmed a comb structure. In the same paper the authors qualitatively explained the comb generation as a consequence of a Kerr-like self-phase modulation occurring in mismatched SHG.
In a later article, the same group extended the spectral analysis to the MIR region around the idler frequency~\cite{Ulvila:2014bx}. In this case the spectral extension of the MIR comb was reduced with respect to that of the NIR signal comb, clearly limited by the phase-matching bandwidth of the DFG process between pump and signal photons. Finally, in a very recent work Ulvila et al.~\cite{Ulvila:2015ia} demonstrated that the spectral quality of the comb could be further improved by modulating the parametric gain via pump laser intensity, and verified the mode-spacing uniformity of the comb at the Hertz level. 
Incidentally, we notice that their original system, apart  from the OPO, can be seen as an internally pumped, phase-mismatched SHG cavity for the signal frequency, similar to the SHG systems we describe below.

\begin{figure}[t]
\begin{center}
\includegraphics*[bbllx=0bp,bblly=0bp,bburx=570bp,bbury=100bp,width=75mm]{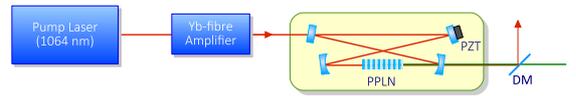}
\caption{Schematic view of the experimental setup. An amplified 1064 nm laser pumps a bow-tie nonlinear cavity with a periodically poled crystal inside. The output radiation is frequency separated and sent to the respective diagnostics, not shown (see text for details).}
\label{fig:setup}
\end{center}
\end{figure}

\subsection{Lithium niobate SHG cavity}
Recently, we demonstrated frequency comb generation in a continuously-pumped cavity-enhanced SHG system, where multiple, cascaded $\chi^{(2)}$ nonlinear processes enable the onset of a broadband comb emission, both around the fundamental pump frequency and its second harmonic~\cite{Ricciardi:2015bw} . We observed different regimes of generation, depending on the phase-matching condition for second-harmonic generation, and, in particular, we showed that phase mismatched SHG is not essential for the appearance of an optical frequency comb in quadratic nonlinear media.

The SHG system was based on a 15-mm-long periodically poled, LiNbO$_3$ crystal, with a grating period of $\Lambda$ = 6.96~$\mu$m placed in a traveling-wave optical cavity (Fig.~\ref{fig:setup}). The pump source was a cw narrow-linewidth Nd:YAG laser, ($\lambda_0$=1064.45~nm), amplified by an Yb:fibre amplifier up to 9~W. The cavity FSR was 493.00(1)~MHz, with a cold cavity resonance full width at half maximum of 3~MHz (finesse: 160; $Q$-factor: 10$^8$). The SHG process for the fundamental wavelength was phase-matched at a crystal temperature $T_0 = 39.5^\circ$C. The crystal temperature was actively stabilized by a Peltier element driven by an electronic servo control. A PDH scheme was implemented to lock the SHG cavity to the pump frequency. Fundamental and harmonic light beams exiting the cavity were separated by a dichroic mirror and sent to different diagnostic systems, consisting of an optical spectrum analyzer (OSA) and radio frequency (RF) spectrum analyzers for the detection of beat notes. A confocal Fabry--P\'erot interferometer was used for optical spectral analysis in the green part of the visible range, not covered by the OSA. Here, we show only IR spectra around the fundamental frequency, nevertheless corresponding optical spectra were observed in the harmonic region. 

\begin{figure}[t!]
\begin{center}
\includegraphics*[bbllx=0bp,bblly=0bp,bburx=520bp,bbury=440bp,width=75mm]{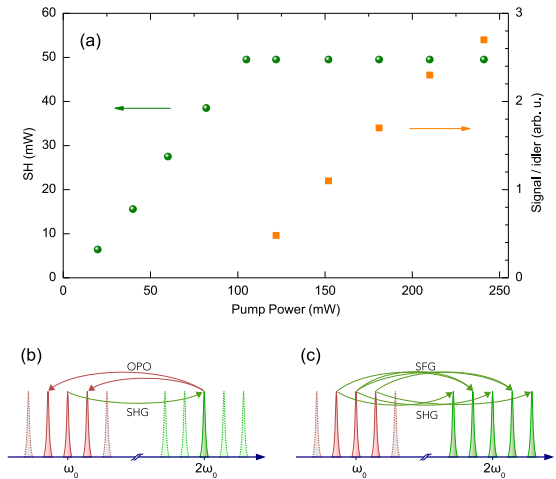}
\caption{(a) Second harmonic and intracavity parametric power as a function of the pump power. When the pump power exceeds 100~mW the SH power clamps and new parametric sidebands appear with increasing power. (b) Schematic representation of second-harmonic generation with a cascading OPO giving rise to a signal-idler pair around the fundamental frequency, oscillating in symmetrically displaced cavity resonances. (c) The multiple subharmonic components, in turn, lead to thresholdless, multiple second-harmonic and sum-frequency mixing processes, eventually culminating in a double comb emission.}
\label{fig:clamping}
\end{center}
\end{figure}

\begin{figure}[t]
\begin{center}
\includegraphics*[bbllx=30bp,bblly=30bp,bburx=530bp,bbury=710bp,width=75mm]{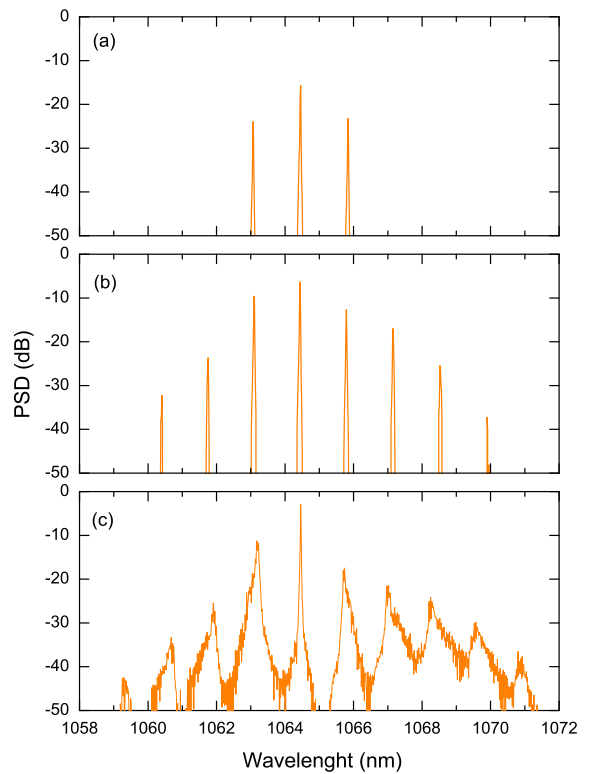}
\caption{Optical spectra around the fundamental frequency. (a) Above the threshold for the internally pumped OPO a couple of parametric waves appears around the fundamental frequency. (b) Increasing the pump power, new cascaded processes originate new sidebands equally spaced by  a multiple of the cavity FSR. (c) Further increase of power leads to formation of secondary frequency combs around each primary sideband. The secondary comb lines are spaced by the cavity FSR, as verified by radio frequency spectral analysis of the output radiation.}
\label{fig:LN-QPM}
\end{center}
\end{figure}

When the crystal was phase-matched for SHG, we observed a first regime of pure harmonic generation [Fig.~\ref{fig:clamping}(a)]. As the laser power exceeded a threshold value of about 100~mW, the second-harmonic power clamped at a constant value and a signal-idler pair started to oscillate symmetrically around the fundamental frequency [Fig.~\ref{fig:LN-QPM}(a)]. Increasing the pump power above the OPO threshold, more sidebands appeared around the fundamental mode, forming a comb whose teeth are equally spaced by an integer multiple $\Delta$$\nu$ of the cavity FSR [Fig.~\ref{fig:LN-QPM}(b)]. 
Comb teeth separation and its uniformity were measured to be an integer multiple of the cavity FSR with a relative uncertainty of $10^{-6}$~\cite{Ricciardi:2015bw}.

When the pump power was further increased, typically $P_{in} > $ 5~W, secondary frequency combs around the primary comb teeth appeared, as shown in Fig.~\ref{fig:LN-QPM}(c), similarly to hierarchical comb formation observed in Kerr-combs~\cite{Papp:2013uu,Herr:2014ip}.
In this case, the RF spectral analysis of the IR and green output light shows intermodal beat notes at the cavity FSR frequency, confirming the regular discreteness of a comb emission with one FRS teeth spacing. 

\begin{figure}[t]
\begin{center}
\includegraphics*[bbllx=30bp,bblly=30bp,bburx=470bp,bbury=770bp,width=75mm]{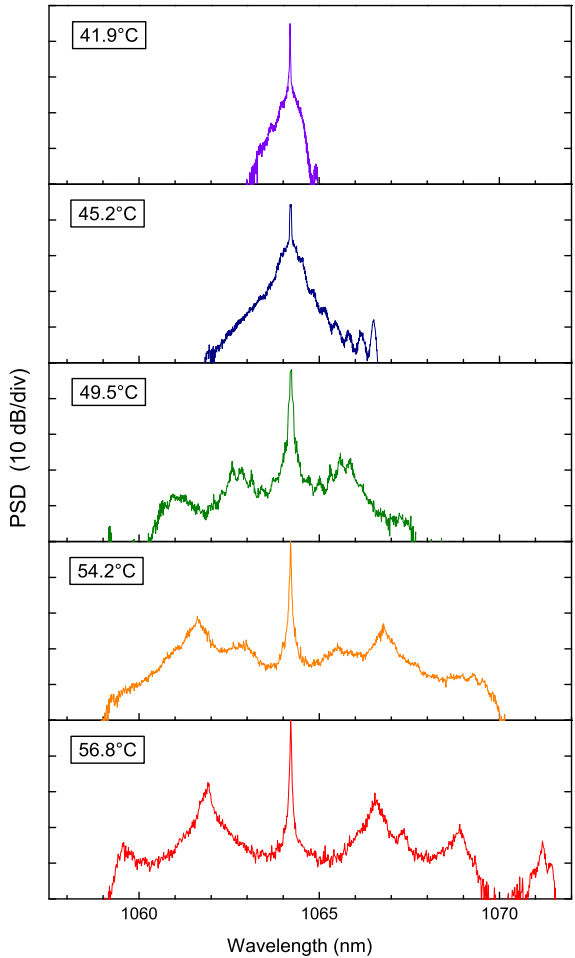}
\caption{Experimental spectra for phase mismatched SHG. From above to below, the crystal temperature has been progressively increased, moving farthe the phase matching condition for the initial SHG. All the spectra have been acquired with the maximum available pump power of 9~W. Power calibration gave an average power per teeth of the order of microwatts.}
\label{fig:outQPM}
\end{center}
\end{figure}

\begin{figure}[t]
\begin{center}
\includegraphics*[bbllx=40bp,bblly=40bp,bburx=700bp,bbury=400bp,width=75mm]{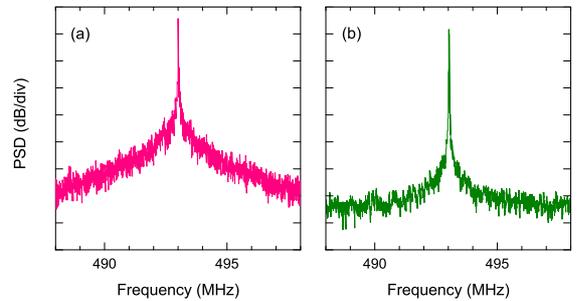}
\caption{RF intermodal beat notes at 493~MHz, corresponding to the cavity FSR, for (a) the IR and (b) the green combs.}
\label{fig:beatnotes}
\end{center}
\end{figure}

Increasing the crystal temperature, the original SHG process became positive phase mismatched and the off-phase matched pump frequency acted as a seed for a 1-FSR-spaced comb.
 In our experiment, we changed the crystal temperature exploring SHG wave-vector mismatches up to 8$\pi$, limited by the working range of the Peltier servo control. Because of the strong photothermal effects, passive thermal locking was exploited for keeping the cavity nearly resonant with the pump. Figure~\ref{fig:outQPM} shows optical spectra of the comb emission for different crystal temperatures at the maximum pump power of 9~W. The comb bandwidth increases with the mismatch temperature, up to more than 10~nm ($\approx$ 5000 teeth). 
The emitted power per mode was estimated to be of the order of microwatts. 
Again, when the IR comb emerged around the fundamental wave, a visible comb was also present around its second harmonic. 
Figure~\ref{fig:beatnotes} shows the beat notes at 493~MHz for the IR and visible combs, confirming the minimal teeth spacing of one FSR. The  resolution-limited, narrow beat notes reveal a high degree of correlation between comb teeth.
We notice that, in the previous case of  phase matched SHG for the fundamental frequency,  the sidebands of the primary comb of Fig.~\ref{fig:LN-QPM}(b) are indeed phase mismatched for the respective SHG process; hence, the appearence of secondary combs around each primary sideband is similar to the direct formation of a closely spaced comb for a phase mismatched pump. 

\subsection{Lithium tantalate SHG cavity}
Despite the sightly lower nonlinear coefficient, with respect to lithium niobate,  lithium tantalate, LiTaO$_3$, shows a wider transparency range 0.28–5.5~$\mu$m, high thermal conductivity, high photorefractive damage threshold, and high resistance to photochromic effects like green-induced infrared absorption. Indeed, stable, single-frequency and high power green sources based on SHG in lithium tantalate crystals have  been realized both in single pass~\cite{Samanta:2009ab,Sinha:2008ab} and intracavity configurations, reaching conversion efficiencies up to 75\%~\cite{Ricciardi:2010ab}. 
Therefore, we tested $\chi^{(2)}$ comb generation in a lithium tantalate based SHG system.

\begin{figure}[t]
\begin{center}
\includegraphics*[bbllx=30bp,bblly=40bp,bburx=705bp,bbury=450bp,width=75mm]{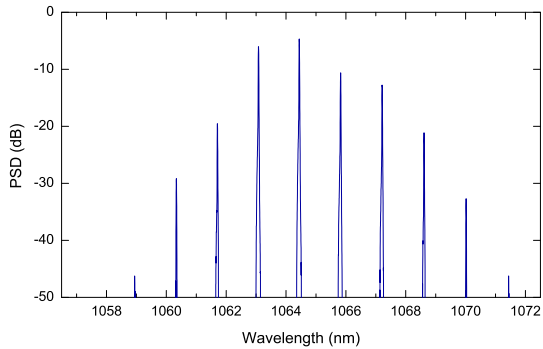}
\caption{Experimental OSA spectrum around the fundamental mode for quasi-phase-matched SHG in lithium tantalate crystal at 4.5~W of input pump power.}
\label{fig:CombLT_PM}
\end{center}
\end{figure}

The experimental set-up was the same used for $\chi^{(2)}$ comb generation in lithium niobate, except for the nonlinear crystal and the corresponding phase matching temperature. We used a 15-mm-long sample of periodically poled 1\%-MgO-doped stoichiometric lithium tantalate, MgO:LiTaO$_3$, with a grating period of 7.97~$\mu$m.
The SH cavity was locked to the stabilized pump frequency by the use of the PDH scheme, and stable lock was achieved for pump powers up to 6.3~W, beyond which active locking is prevented by strong photothermal effects. We investigated both the phase matching and positive phase mismatched regime.  
At a phase-matching temperature of 35.5$^\circ$C, similarly to what we observed with lithium niobate, the SH power increased with the pump power with an efficiency of about~50\%. For pump powers higher than 300~mW, the second harmonic power clamped and the cascaded OPO started oscillating, generating s-i couples around the fundamental. 
Figure~\ref{fig:CombLT_PM} shows the OSA spectrum around the fundamental mode for input pump power of 4.5~W, when equally spaced parametric sidebands oscillate around the fundamental.
In comparison with lithium niobate, due to the intrinsic lower nonlinear coefficient, lithium tantalate results in a higher threshold for the cascaded OPO. 
For the same reason we did not observed a clear evidence of secondary comb formation as in lithium niobate.

\begin{figure}[t]
\begin{center}
\includegraphics*[bbllx=30bp,bblly=40bp,bburx=705bp,bbury=440bp,width=75mm]{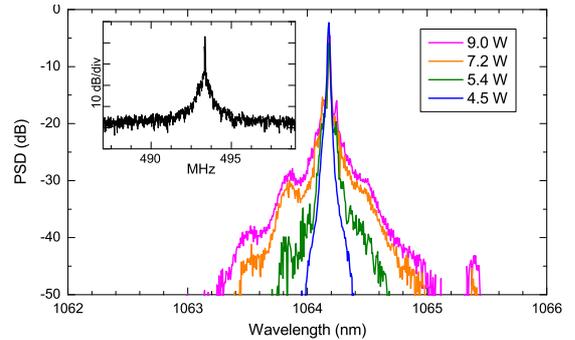}
\caption{Experimental spectra for off-phase-matched SHG in lithium tantalate crystal at a temperature of $\textrm{39.0}\:^\circ$C for different input pump powers. The inset shows the beat note of the infrared radiation at 493~MHz.}
\label{fig:LT_NPM}
\end{center}
\end{figure}

We also explored the positive phase-mismatching condition: Fig.~\ref{fig:LT_NPM} shows an example of comb generation in off-phase-matched SHG for different input pump powers from 4.5 up to 9~W, under passive thermal locking. As we can see, the generated comb around the fundamental frequency shows a trend to broadening as a function of the input pump power. 

\subsection{Truncated three-wave dynamic model}
\label{TTW}
In view of the observed emission regimes and of previous modeling of SHG with internally pumped OPO~\cite{Schiller:1997dp}, we developed a simple mathematical model based on truncated three-wave dynamic equations for the resonant subharmonic fields.
In our system, the SHG cavity resonates for the subharmonic range of frequencies around the fundamental pump frequency, while the harmonic fields leave the cavity after generation in the nonlinear medium. 
Compared to the model of Ref.~\cite{Schiller:1997dp}, which considered only the two processes of frequency doubling and cascaded degenerate OPO, we included a more complete set of second-order processes which start after the cascaded OPO, namely  generation of the signal and idler second harmonic, and sum frequency of signal (idler) and fundamental, as shown in Fig.~\ref{fig:clamping}(b) and (c). These processes lead to a complete, closed three-wave dynamical model for the resonant fields. Once these processes are considered, the derived dynamic equations are still limited to three subharmonic fields, but with new interaction terms which are relevant for a deeper comprehension. In particular, the sum frequency mixing processes are the leading mechanism of nonlinear losses, at least near the threshold and determine the position of the first oscillating signal-idler pair, as shown in Sec.~\ref{S:stability}.

Based on this picture, starting from coupled mode equations for the extended set of basic $\chi^{(2)}$ processes, we perturbatively derived an elemental nontrivial system of dynamic equations for the sole resonating sub-harmonic fields, while the harmonic fields could be considered fast variables whose dynamics is slaved to the sub-harmonic one, leading to~\cite{Ricciardi:2015bw}:
\begin{subequations}
\begin{align}
\dot{A}_0 =& -(\gamma + i \Delta_0) A_0 
- 2 g_0 \eta_{0si} A_0^* A_\text{i} A_\text{s} 
- g_0 (\eta_{00} |A_0|^2
\nonumber\\&
+2\eta_\text{0s} |A_\text{s}|^2+2\eta_\text{0i} |A_\text{i}|^2) A_0  
+F_\text{in} 
\label{eq:1-a}
\\
\dot{A}_\text{s} =& -(\gamma + i \Delta_\text{s}) \, A_\text{s}
- g_0 \eta_\text{00i} \, A_0^2 A^*_\text{i} 
- g_0 (2 \eta_\text{s0} |A_0|^2
\nonumber\\&
+\eta_\text{ss} |A_\text{s}|^2 + 2\eta_\text{si} |A_\text{i}|^2) A_\text{s}  
\label{eq:1-b}
\\
\dot{A}_\text{i} =& -(\gamma + i \Delta_\text{i}) \, A_\text{i}  
- g_0  \eta_\text{00s} \, A_0^2 A^*_\text{s}
- g_0 (2 \eta_\text{i0} |A_0|^2 
\nonumber\\&
+ 2 \eta_\text{is} |A_\text{s}|^2 + \eta_\text{ii} |A_\text{i}|^2) A_\text{i} 
 \,.
\label{eq:1-c}
\end{align}
\label{eq:1}
\end{subequations}
Here subscripts 0, ‘s’ and ‘i’ indicate fundamental, signal and idler modes, respectively. The A’s are the normalized electric field amplitudes whose squared modulus $|A|^2$ is a photon number; $F_{in}$ is the pump amplitude coupled into the cavity; $\gamma$ is the cavity decay constant, assumed to be the same for the three fields; the $\Delta$’s are the cavity detunings of the respective modes; the $\eta$’s are complex nonlinear coupling constants, depending on the wave-vector mismatches of the considered second-order processes; and $g_0$ is a common gain factor depending on the crystal length and the second-order coupling strength (see Ref.~\cite{Ricciardi:2015bw} or Sec.~\ref{sec:modal-eq}, for explicit definitions).

Despite the different underlying physical mechanism, Eqs.~(\ref{eq:1}) reveal a  striking resemblance to FWM-based models for Kerr combs generation in microresonators  ~\cite{Chembo:2010ii,Hansson:2013jy}. The constants $\eta$’s are effective third-order complex susceptibilities, where the real and imaginary parts give the ‘absorption’ and ‘dispersion’ components, respectively. Focusing on  the interaction terms $|A_l|^2 A_l$, and $|A_m|^2A_l|$ (with $l,m \in \left\{0,s,i\right\}$ and $l \neq m$), their imaginary parts correspond, respectively, to self- and cross-phase modulation terms, producing an effective change of the refractive index, which locally  compensate the unequal spacing of the cavity modes due to the group velocity dispersion (GVD) of the material, so that the cavity modes become locally equidistant. On the other hand, the corresponding real parts represent quadratic nonlinear losses for the resonant fields and indeed determine the frequency distance from the fundamental mode at which a s-i pair oscillates.

\section{Generalized modal equations}
\label{sec:modal-eq}
In this section we present a heuristical generalization of the truncated three-wave model of Eqs.~(\ref{eq:1}) to the case of a $N$-teeth $\chi^{(2)}$ comb, thus describing the dynamics of a comb well above the cascaded OPO threshold. 
To this end, we first derive the reduced dynamic equations of the resonant subharmonic fields for a bichromatically pumped nonlinear SHG cavity,  showing as they mimick a nondegenerate FWM third-order process where the initial pump waves at $\omega_2$ and $\omega_3$ give rise to two new waves at $\omega_1$ and $\omega_4$, so as to satisfy energy conservation, $\omega_2+\omega_3=\omega_1+ \omega_4$.
As shown for Kerr combs, bichromatic pumping is in fact a practical scheme for efficient, thresholdless frequency comb generation and comb teeth separation control~\cite{Strekalov:2009ee,Hansson:2014cw}; furthermore, it can be useful for accurate frequency stabilization of the comb emission as required for metrological applications. 
Moreover, the presence of two initial subharmonic fields is a helpful simplified picture of what happens after the internally pumped OPO starts and each sideband interacts with the pump field and with other sidebands.

\subsection{Bichromatically pumped SHG cavity}
\begin{figure}[t!]
\begin{center}
\includegraphics*[bbllx=0bp,bblly=0bp,bburx=250bp,bbury=300bp,width=75mm]{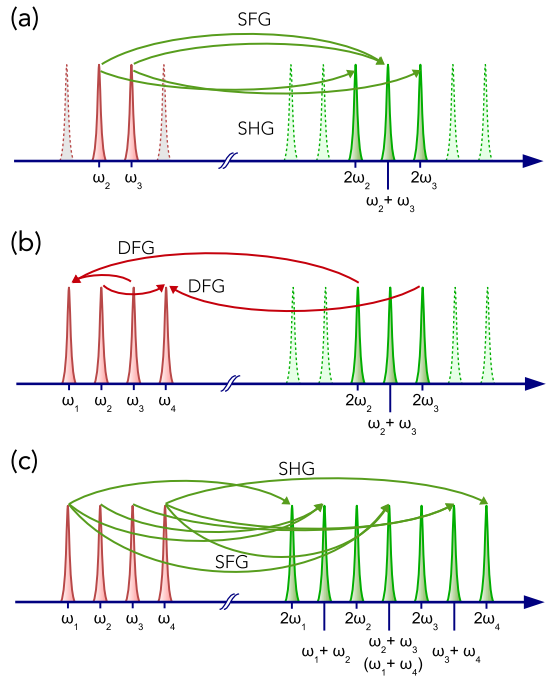}

\caption{Schematic representation of cascaded quadratic processes mimicking nondegenerate FWM in the case of a bichromatically pumped cavity: (a) the initial presence of two input sub-harmonic modes, $\omega_2$ and $\omega_3$, enables the generation of three harmonic fields by SHG and SFG; (b) DFGs between the second harmonics and the original sub-harmonic fields result in two new symmetrically placed output sub-harmonic fields, $\omega_1$ and $\omega_4$, which successively (c) generate other harmonic fields. From the point of view of the sub-harmonic fields the process is equivalent to a $\chi^{(3)}$ non degenerate FWM.}
\label{fig:ndFWM}
\end{center}
\end{figure}

Figure~\ref{fig:ndFWM} schematically illustrates all the cascaded $\chi^{(2)}$ processes which limit the involved sub-harmonic fields to four.
Initially, the waves at $\omega_2$ and $\omega_3=\omega_2+\Delta\omega$ propagating in the quadratic medium can generate their respective second harmonic waves, $2\omega_2$ and $2\omega_3$, along with the sum frequency, $\omega_{23}=\omega_2+\omega_3$. We note that the frequency  separation between two nearby harmonic wave equals the initial spacing $\Delta\omega$. 
Then, each second harmonic wave interacts with the opposite sub-harmonic leading, through DFG, to new waves at $\omega_1=2\omega_2-\omega_3=\omega_2-\Delta\omega$ and $\omega_4=2\omega_3-\omega_2=\omega_3+\Delta\omega$. Finally, others SH and sum frequencies are produced, uniformly filling the harmonic range.
In fact, the number of cascaded processes rapidly grows, generating new sub-harmonic and harmonic fields, all equally spaced by $\Delta\omega$, but here we limit our analysis to the first four sub-harmonic fields and their seven harmonic combinations.

We adopt the same notation and field amplitude normalization of Ref.~\cite{Ricciardi:2015bw}, according to which $|A_j|^2$ gives the photon number of field $A_j$. We omit to explicitly write down the eleven coupled mode equations, as it is straigthforward but lengthy.  
For sake of simplicity, in the following we assume a travelling wave cavity, completely filled with the nonlinear medium so that the nonlinear interaction length and  the cavity geometrical path coincide, i.e., $L_\text{cav}=L$.  
It is straightforward to generalize to the case of a partially filled cavity.

We notice that each second-order process depicted in Fig.~\ref{fig:ndFWM} involves two sub-harmonic, $\omega_j$ and $\omega_k$ (possibly degenerate, i.e.,  $j=k$) and one harmonic field, $\omega_j+\omega_k$. 
Therefore, for each process, we identify the matching wave vector as,
\begin{equation}
\xi_{jk} = k_{\omega_{j}+\omega_{k}}- k_{\omega_{j}} -k_{\omega_{k}} \, ,
\label{eq:wavevector}
\end{equation}
which is invariant under exchange of the indices, i.e., $\xi_{jk} =\xi_{kj} $. 
As it will be soon clear, this notation will facilitate the generalization of the dynamic equations for a generic number of sub-harmonic fields.
Usually, for a given crystal condition and initial field frequency, only one of the processes can be perfectly phase matched and, more generally, none of them is. 

The coupled mode equations can be perturbatively solved by successive integrations, along the lines of the procedure outlined in Ref.~\cite{Ricciardi:2015bw}, including the effect of the optical cavity, finally yielding the following dynamic equations: 
\begin{subequations}
\begin{align}
\dot{A}_2 =& -(\gamma_2 + i \Delta_2) \, A_2 
-  g_0 \left[ \sum_{\nu=1}^4 (2-\delta_{2\nu}) \, \eta_{2\nu2\nu} |A_{\nu}|^2 \right] A_{2} 
 \nonumber \\
&
- 2 g_0 \, \eta_{2231} \, A_2^* A_3 A_1
- 2 g_0 \, \eta_{2314} \, A_3^* A_1 A_4
 \nonumber \\
&
- g_0 \, \eta_{2433} \, A_4^* A_3^2 
+ F_\text{2}  
\label{eq:FW-a}
\\
\dot{A}_3 =& -(\gamma_3 + i \Delta_3) \, A_3 
-  g_0 \left[ \sum_{\nu=1}^4 (2-\delta_{3\nu}) \, \eta_{3\nu3\nu} |A_{\nu}|^2 \right] A_{3} 
 \nonumber \\
&
- 2 g_0 \, \eta_{3324} \, A_3^* A_2 A_4
- 2 g_0 \, \eta_{3214} \, A_2^* A_1 A_4
 \nonumber \\
&
- g_0 \, \eta_{3122} \, A_1^* A_2^2 
+ F_\text{3}  
\label{eq:FW-b}
\\
\dot{A}_1 =& -(\gamma_1 + i \Delta_1) \, A_1 
-  g_0 \left[ \sum_{\nu=1}^4 (2-\delta_{1\nu}) \, \eta_{1\nu1\nu} |A_{\nu}|^2 \right] A_{1} 
 \nonumber \\
&
- 2 g_0 \, \eta_{1423} \, A_4^* A_2 A_3
- g_0 \, \eta_{1322} \, A_3^* A_2^2 
\label{eq:FW-c}
\\
\dot{A}_4 =& -(\gamma_4 + i \Delta_4) \, A_4 
-  g_0 \left[ \sum_{\nu=1}^4 (2-\delta_{4\nu}) \, \eta_{4\nu4\nu} |A_{\nu}|^2 \right] A_{4} 
 \nonumber \\
&
- 2 g_0 \, \eta_{4123} \, A_1^* A_2 A_3
- g_0 \, \eta_{4233} \, A_2^* A_3^2   \, ,
\label{eq:FW-d}
\end{align}
\label{eq:FW}
\end{subequations}
where $g_0=(\kappa L)^2/4\tau$, with $\kappa=d/c\sqrt{2\omega_0^3/n_1^2 \, n_2}$ the quadratic nonlinear coupling constant, $L$ the interaction length, and $\tau$ the cavity round-trip time; $\delta_{\mu\nu}$ is the Kronecker's delta. $F_j=\sqrt{2\gamma_j/\tau}A_j^\text{in}$ is the round-trip coupled field amplitude,  with $A_j^\text{in}$ being the input field amplitude around the mode $j=2,3$. The complex coupling constants are given by
\begin{equation}
\eta_{\mu\nu\rho\sigma} = I (\xi_{\mu\nu},\xi_{\rho\sigma}) =\frac{2}{L^2} \int_{0}^{L} 
e^{-i \xi_{\mu\nu}z} \, \frac{e^{i\xi_{\rho\sigma} z}-1}{i \xi_{\rho\sigma}} \; \text{d}z \, .
\label{eq:eta}
\end{equation}

\begin{figure}[t!]
\begin{center}
\includegraphics*[bbllx=0bp,bblly=0bp,bburx=310bp,bbury=200bp,width=75mm]{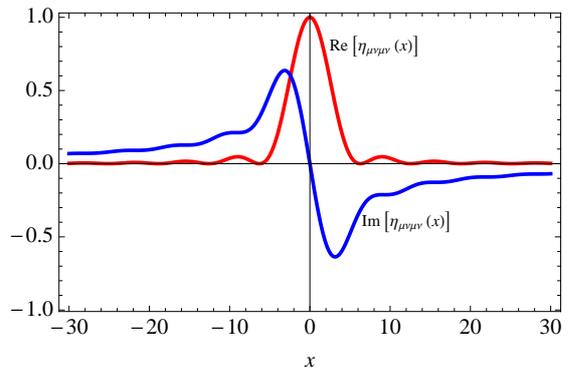}
\caption{Plot of Eqs.~(\ref{eq:eta_abab1}) and (\ref{eq:eta_abab2}), real and imaginary part of the coupling constant $\eta_{\mu\nu\mu\nu}$.}
\label{fig:Eta_abab}
\end{center}
\end{figure}

Figure \ref{fig:Eta_abab} shows the absolute value of the two-variable function $I (x,y)$ and the real and imaginary parts of the index-symmetric, single-variable function $\eta_{\mu\nu\mu\nu}(x)$ which read
\begin{eqnarray}
\text{Re}[\eta_{\mu\nu\mu\nu}(x)] &=& \text{sinc}^2 (x/2)	
\label{eq:eta_abab1}\\
\text{Im}[\eta_{\mu\nu\mu\nu}(x)] &=& \frac{2}{x} ( \text{sinc} (x/2) - 1)  \, .
\label{eq:eta_abab2}
\end{eqnarray}

The coupling constants $\eta$'s, characterizing the strength of the effective third-order interactions, express indeed the effect of a couple of cascaded second-order processes, integrated over the length of the nonlinear medium, according to Eq.~(\ref{eq:eta}). 
They can be generally seen as effective complex susceptibilities, whose real and imaginary parts, differently from the usual definition, are related to absorption and dispersion, respectively.
Furthermore, as a consequence of the invariance of the wave vectors under the exchange of indices, Eq.~(\ref{eq:wavevector}), the coupling constants $\eta$ are invariant under the independent exchange of the first or last pair of indices, i.e., $\eta_{\mu\nu\rho\sigma}=\eta_{\nu\mu\rho\sigma}=\eta_{\nu\mu\sigma\rho}$.

From the solution of the simple case of effective non degenerate FWM, Eqs. (\ref{eq:FW}), we can heuristically derive a general expression for the dynamic equations for any number of interacting fields, yielding for each field $A_{\mu}$, nearly resonant with the  $\mu$-th cavity mode,
\begin{equation}
\dot{A}_{\mu} = -(\gamma_{\mu} + i \Delta_{\mu}) \, A_{\mu}
	- g_0 \!\!\!\!\!\! \sum_{\substack{{\rho,\sigma}\\{\nu=\rho+\sigma-\mu}}} \!\!\!\!\!\! \eta_{\mu\nu\rho\sigma} \, A_{\nu}^* A_{\rho} A_{\sigma}  + F_\mu \, ,
\label{eq:FWM-gen}
\end{equation}
where the summation over the indices $\rho$ and $\sigma$ can be extended over all the cavity resonant modes, and $F_\mu$ is the cavity coupled amplitude of a input driving field $A_\mu^\text{in}$, as defined before. 
The constraint over $\nu$ is dictated by the energy conservation condition.

\subsection{Bistability}

\begin{figure*}[t!]
\begin{center}
\includegraphics*[bbllx=0bp,bblly=0bp,bburx=305bp,bbury=200bp,width=75mm]{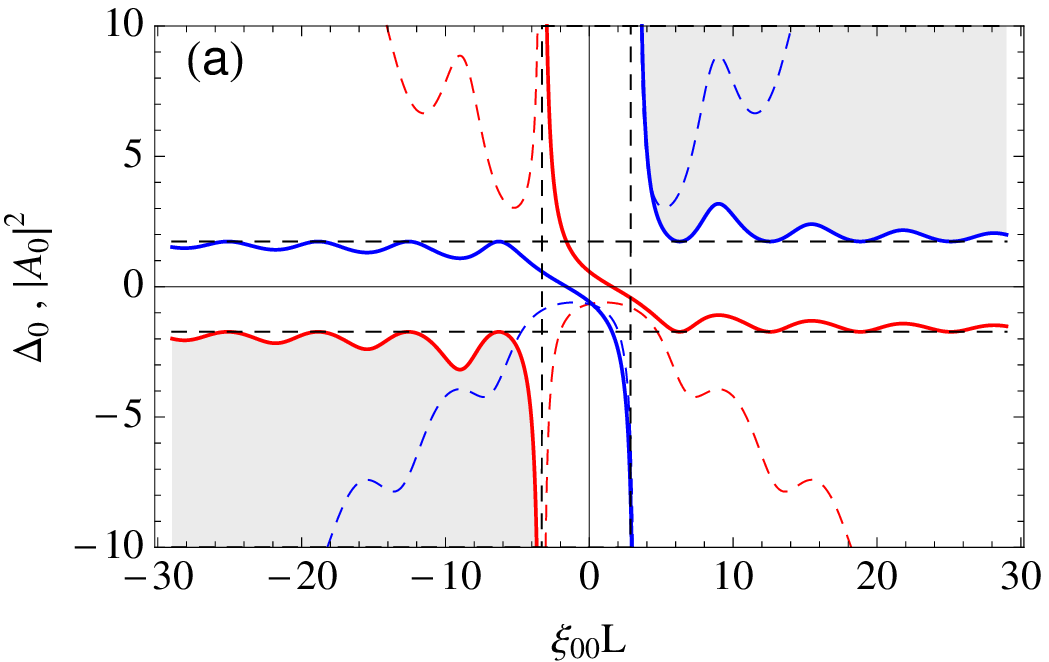}
\includegraphics*[bbllx=0bp,bblly=0bp,bburx=320bp,bbury=200bp,width=75mm]{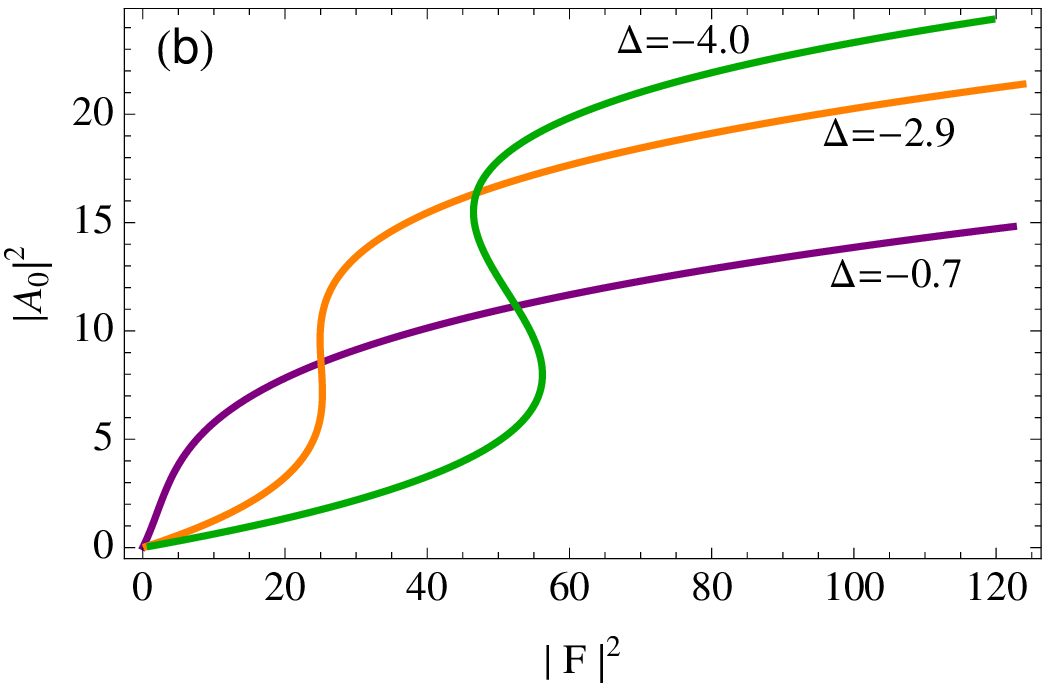}
\caption{(a) Stability diagram of stationary solutions of Eq.~(\ref{eq:cubic}) as a function of the detuning $\Delta$ and the matching parameter $\xi_{00}L$, assuming $g_0=1$ and $\gamma_0=1$. Solid lines represent the critical detunings, $\Delta_-$ (red) and $\Delta_+$ (blue), while the dashed curves are the corresponding intracavity powers $x_\pm$, plotted against the matching parameter $\xi_{00}L$. Physically allowable solution requires $x_\pm>0$, thus bistability occurs for $|\xi_{00}L| \gtrsim 3.29$ (black vertical dashed lines). The shaded regions identify parameters values for which bistability occurs.
The horizontal dashed lines correspond to $\Delta_0=\pm\sqrt{3}$, which is the critical detuning when $\xi_{00}L$ is an integer multiple of $2\pi$ and there is no nonlinear loss due to second-harmonic generation, thus the nonlinear cavity acts as a purely dispersive Kerr cavity. 
(b) Intracavity power $|A_0|^2$ as a function of injected power $|F|^2$, for $\xi_{00}L=10$ and $\Delta=-4.0$, $-2.9$, and $-0.7$, showing the transition from single-valued solution to three-valued bistable solution. The negative slope segment of the three-valued solution is always unstable.}
\label{fig:bistability}
\end{center}
\end{figure*}

In the last part of this section we consider the case where only the fundamental mode $A_0$ is excited by a single-mode external pump, corresponding to the simple SHG system below the threshold for the internally pumped OPO. Apparently trivial, this configuration exhibits bistable regimes analogously to a Kerr cavity and was investigated theoretically and experimentally~\cite{White:1996ts}. 
Here, we resume the main result concerning its stability analysis.
 
For this case, Eq.~(\ref{eq:FWM-gen}) reduces to
\begin{equation}
\dot{A}_0 = -(\gamma_0 + i \Delta_0) \, A_0 - g_0 \eta_{0000} |A_0|^2  \, A_0  + F_0  \, .
\label{eq:bistab1}
\end{equation}
As the coupling constant $\eta_{0000}$ is in general a complex number, the nonlinear term, proportional to the intracavity power $|A_0|^2$, gives: a nonlinear loss, proportional to $\text{Re}(\eta_{0000})$, due to fundamental power converted into second harmonic and leaving the cavity; and a nonlinear detuning, proportional to $\text{Im}( \eta_{0000})$. 
The latter, which occurs for phase mismatched SHG, i.e., $\xi_{00}\ne0$, may lead to a bistable regime~\cite{White:1996ts}. 
Actually, there is a detuning threshold and a power threshold which mark the onset of bistability.
To determine these threshold values we consider the steady state solution of Eq.~(\ref{eq:bistab1}) which expresses the relationship between the driving power, $y=|F|^2$, and intracavity power, $x=|A_0|^2$, as
\begin{equation}
y=\left(R^2+J^2\right) x^3+ 2(\gamma_0  R+\Delta_0  J)  x^2+\left(\gamma_0^2+\Delta_0^2\right) x \, ,
\label{eq:cubic}
\end{equation}
where $R=g_0 \text{Re}[\eta_{0000}(\xi_{00}L)]$ and $J=g_0 \text{Im}[\eta_{0000}(\xi_{00}L)]$.
According to the cubic Eq.~(\ref{eq:cubic}), for a given input power $y$ there could be one or three solutions for the intracavity power $x$, provided that the physically meaningful condition $x>0$ is satisfied. 
In particular, bistability occurs when the right-hand-side of Eq.~(\ref{eq:cubic}) have local extrema, e.g., if there are values of $x$ for which the derivative $\partial y/\partial x = 0$, or 
\begin{equation}
3 (J^2 + R^2) x^2  +  4  ( R \gamma_0 + J \Delta_0) x + \gamma_0^2 + \Delta_0^2 =0
\label{eq:derivata}
\end{equation}
This equation has real, nonnegative solutions for detunings 
\begin{equation}
\begin{array}{l} \Delta_0\le\Delta_- \quad \text{if} \quad J> \sqrt{3}\, R  \, ,
\\ 
\Delta_0\ge\Delta_+ \quad \text{if} \quad  -J> \sqrt{3}\, R  \, ,
\end{array}
\label{eq:delta-cond}
\end{equation} 
 with the critical detunings
\begin{equation*}
\Delta _{\pm }=\frac{-4 J R \pm \sqrt{3} \left(J^2+R^2\right) }{J^2 - 3 \, R^2} \; \gamma_0 \, .
\end{equation*} 
The corresponding intracavity power threshold is obtained by substituting $\Delta _{\pm}$ in Eq.~(\ref{eq:derivata}) and solving for $x$, yielding
\begin{equation*}
x_\pm= \frac{ 2\gamma_0}{\sqrt{3}} \; \frac{1}{ |J| \mp \sqrt{3}\, R}  \, .
\end{equation*}
The conditions of Eq.~(\ref{eq:delta-cond}) on $R$ and $J$ imply a condition on the matching wave vector, $|\xi_{00}L| \gtrsim 3.29$. 
Finally, from Eq.~(\ref{eq:cubic}) the input threshold power is,
\begin{equation*}
y_\pm= \frac{8 \gamma_0^3}{ 3\sqrt{3}} \; \frac{1}{( |J| \mp \sqrt{3} \, R)^3} \, .
\end{equation*}

Figure \ref{fig:bistability}(a) shows the bistability map as a function of the detuning $\Delta_0$ and the matching wave vector  $\xi_{00}$, having assumed $g_0=1$ and $\gamma_0=1$. 
The shaded regions identify parameters values for which bistable solutions are possible. For matching wave vectors such that $|\xi_{00}L| \lesssim 3.29$, steady state solutions for the intracavity power are always single-valued. 
Moreover, when $\xi_{00}L$ is an integer multiple of $2\pi$, then $\text{Re}[\eta_{0000}(\pi)]=0$ and, as a consequence, there is no net second-harmonic generation and the cavity acts as a purely dispersive Kerr cavity~\cite{Khalaidovski:2009ga}. 
Figure \ref{fig:bistability}(b) shows the steady state solution for the intracavity power vs. the input power, Eq.~(\ref{eq:cubic}), for $\xi_{00}L=10$ and $\Delta=-0.7$, $-2.9$, and $-4.0$, respectively. Notice that from the anti-symmetric features of the stability diagram the same solutions holds by simultaneously changing the sign of the matching wave vector and the detuning.  

\section{Subharmonic pumped OPO: linear stability analysis}
\label{S:stability}

We specialize Eq.~(\ref{eq:FWM-gen}) to the single-pump case, looking for the condition for the appearance of a pair of frequency-symmetric parametric modes.
To this purpose, we assume that each field is nearly resonant with a cavity mode whose position is defined by an integer mode number $\mu$ with respect to a central mode $\varpi_0$, nearly resonant  for the pump, thus yielding the same dynamic equations of Eqs.(\ref{eq:1}), with a more general, compact notation,
\begin{subequations}
\begin{align}
\dot{A}_0 =& 
\!-\!(\gamma_0 + i \Delta_0) \, A_0 
\!- \!2 g_0 \, \eta_{00\mu\bar{\mu}} A_0^* A_{\mu} A_{\bar{\mu}} 
\!-\! g_0 (\eta_{0000} |A_0|^2
\nonumber \\ &
+2\eta_{0\mu0\mu} |A_{\mu}|^2
+2\eta_{0\bar{\mu}0\bar{\mu}} |A_{\bar{\mu}}|^2)  \, A_0  
+ F_0  
\label{eq:d-FWM-a}
\\
\dot{A}_{\mu} =& -(\gamma_{\mu} + i \Delta_{\mu}) \, A_{\mu}  
- g_0 \, \eta_{\mu\bar{\mu}00} \, A_0^2 A^*_{\bar{\mu}} 
- g_0 (2 \eta_{\mu00\mu} |A_0|^2 
\nonumber \\ &
+\eta_{\mu\mu\mu\mu} |A_{\mu}|^2 + 2\eta_{\mu\bar{\mu}\mu\bar{\mu}} |A_{\bar{\mu}}|^2)  \, A_{\mu}
\label{eq:d-FWM-b}
\\
\dot{A}_{\bar{\mu}} =&
-(\gamma_{\bar{\mu}} + i \Delta_{\bar{\mu}}) \, A_{\bar{\mu}} 
- g_0 \eta_{\bar{\mu}\mu00} \, A_0^2 A^*_{\mu}
- g_0 (2 \eta_{\bar{\mu}00\bar{\mu}} |A_0|^2 
\nonumber \\ &
+ 2 \eta_{\bar{\mu}\mu\bar{\mu}\mu} |A_{\mu}|^2 
+ \eta_{\bar{\mu}\bar{\mu}\bar{\mu}\bar{\mu}} |A_{\bar{\mu}}|^2)  \, A_{\bar{\mu}} \, ,
\label{eq:d-FWM-c}
\end{align}
\label{eq:d-FWM}
\end{subequations}
where $\bar{\mu}=-\mu$, for the sake of notation simplicity. 

For a dispersive cavity the frequency position of cavity modes in the vicinity of a central mode $\varpi_0$ can be written as a series expansion with respect to the mode number $\mu$, as
$
\varpi_\mu \simeq \varpi_0 + D_1\, \mu + D_2 \, \mu^2 
$,
where $D_1 = 2\pi c/L_\text{cav} \, n_0$ is the cavity FSR, with $L$ the round-trip optical path of the cavity, and $D_2 \simeq -2 \pi^2 c^3 \beta^{\prime\prime}/L^2 \, n_0^3 = -(c/2 n_0) D_1^2 \beta^{\prime\prime}$ accounts for the group velocity dispersion, $ \beta^{\prime\prime} = d^2k/d\omega^2 |_{\omega_0}$, both calculated at $\omega_0$, with $n_0=n(\omega_0)$.
In fact, in the validity range of the present approximation, the mode number $\mu$ identifies a quasi-symmetric pair ($\mu$-pair) of cavity modes with respect to the fundamental pump mode around $\varpi_0$. 
The detunings between the resonating fields and the nearest cavity modes are, respectively: 
$
\Delta_0 = \omega_0 - \varpi_0 
$; 
$
\Delta_\mu = \omega_\mu - \varpi_\mu= \omega_\mu - \varpi_0 - D_1\, \mu - D_2 \, \mu^2 
$;
$
\Delta_{\bar{\mu}} = \omega_{\bar{\mu}} - \varpi_{\bar{\mu}}= \omega_{\bar{\mu}} - \varpi_0 +D_1\, \mu - D_2 \, \mu^2
$.
We also introduce the variable transformation, 
$
A_0 =a _0 
$; 
$
A_\mu = a_\mu e^{-i (\omega_\mu-\omega_0 - D_1 \mu) t} 
$;
$
A_{\bar{\mu}} = a_{\bar{\mu}} e^{-i (\omega_{\bar{\mu}}-\omega_0 + D_1 \mu) t} \, .
$
After some manipulation, we rewrite Eqs.~(\ref{eq:d-FWM}) as
\begin{subequations}
\begin{align}
\dot{a}_0 =& 
-(\gamma + i \Delta_0) \, a_0 
- 2 g_0 \, \eta_{00\mu\bar{\mu}} a_0^* a_{\mu} a_{\bar{\mu}} 
- g_0 (\eta_{0000} |a_0|^2
\nonumber \\ &
+2\eta_{0\mu0\mu} |a_{\mu}|^2
+2\eta_{0\bar{\mu}0\bar{\mu}} |a_{\bar{\mu}}|^2)  \, a_0  
+ F_0  
\label{eq:d-FWM2-a}
\\
\dot{a}_{\mu} =& 
-[\gamma +i ( \Delta_{0} - D_2 \mu^2)] \, a_{\mu}  
\nonumber \\ &
- g_0 \, \eta_{\mu\bar{\mu}00} \, a_0^2 a^*_{\bar{\mu}} 
- g_0 (2 \eta_{\mu00\mu} |a_0|^2 
\nonumber \\ &
+\eta_{\mu\mu\mu\mu} |a_{\mu}|^2 
+ 2\eta_{\mu\bar{\mu}\mu\bar{\mu}} |a_{\bar{\mu}}|^2)  \, a_{\mu}
\label{eq:d-FWM2-b}
\\
\dot{a}_{\bar{\mu}} =& 
-[\gamma +i ( \Delta_{0} - D_2 \mu^2)] \, a_{\bar{\mu}} 
\nonumber \\ &
- g_0 \eta_{\bar{\mu}\mu00} \, a_0^2 a^*_{\mu}
- g_0 (2 \eta_{\bar{\mu}00\bar{\mu}} |a_0|^2 
\nonumber \\ &
+ 2 \eta_{\bar{\mu}\mu\bar{\mu}\mu} |a_{\mu}|^2 
+ \eta_{\bar{\mu}\bar{\mu}\bar{\mu}\bar{\mu}} |a_{\bar{\mu}}|^2)  \, a_{\bar{\mu}} \, ,
\label{eq:d-FWM2-c}
\end{align}
\label{eq:d-FWM2}
\end{subequations}
where we have assumed $\gamma_{0}=\gamma_{\mu}=\gamma_{\bar{\mu}}=\gamma$.

To predict the conditions for which a $\mu$-pair of parametric fields starts to oscillate we study the stability of the trivial steady state solution of Eqs.~(\ref{eq:d-FWM2}) with zero parametric field amplitudes, $a_{\mu}=a_{\bar{\mu}}=0$. We introduce small perturbations of the zero parametric amplitudes,  $\delta a_{\mu}$ and $\delta a_{\bar{\mu}}$ and substitute in Eq.~(\ref{eq:d-FWM2-b}) and the complex conjugate of  Eq.~(\ref{eq:d-FWM2-c}). Retaining only the terms linear in the perturbations, we obtain the following linear differential equations for small perturbation of the trivial steady state solution in matrix form, as

\begin{equation*}
\left(\begin{array}{c} \delta \dot{a}_\mu \\ \delta \dot{a}_{\bar{\mu}}^* \end{array}\right) 
=
\left(\begin{array}{cc}M_{11} & M_{12} \\ M_{21} & M_{22} \end{array}\right)
\left(\begin{array}{c} \delta a_\mu  \\ \delta a_{\bar{\mu}}^*\end{array}\right)
\end{equation*}
where
\begin{eqnarray*}
M_{11} &=&
-[\gamma +i ( \Delta_{0}-D_2 \mu^2)+ 2 g_0 \eta_{\mu00\mu} |a_0|^2] 
\\
M_{12} &=& M_{21}^* =
-g_0 \eta_{\mu\bar{\mu}00} \, a_0^2
\\
M_{22} &=&
-[\gamma -i( \Delta_{0}-D_2 \mu^2)+ 2 g_0 \eta_{\bar{\mu}00\bar{\mu}}^* |a_0|^2] \, .
\end{eqnarray*}

We notice that, in general, the characteristic polynomial of the matrix $M$ has complex coefficients. The interaction terms remaining after the linearization are $g_0 \eta_{\mu00\mu} |a_0|^2 $ and $g_0 \eta_{\bar{\mu}00\bar{\mu}}^* |a_0|^2$, related to the SFG of the pump with the signal and the idler, respectively, as anticipated in Sec~\ref{TTW}, while $g_0 \eta_{\mu\bar{\mu}00} \, a_0^2$ is associated to the parametric generation in the cascaded OPO~\cite{Ricciardi:2015bw}.

\begin{figure}[t!]
\begin{center}
\includegraphics*[bbllx=30bp,bblly=300bp,bburx=705bp,bbury=700bp,width=75mm]{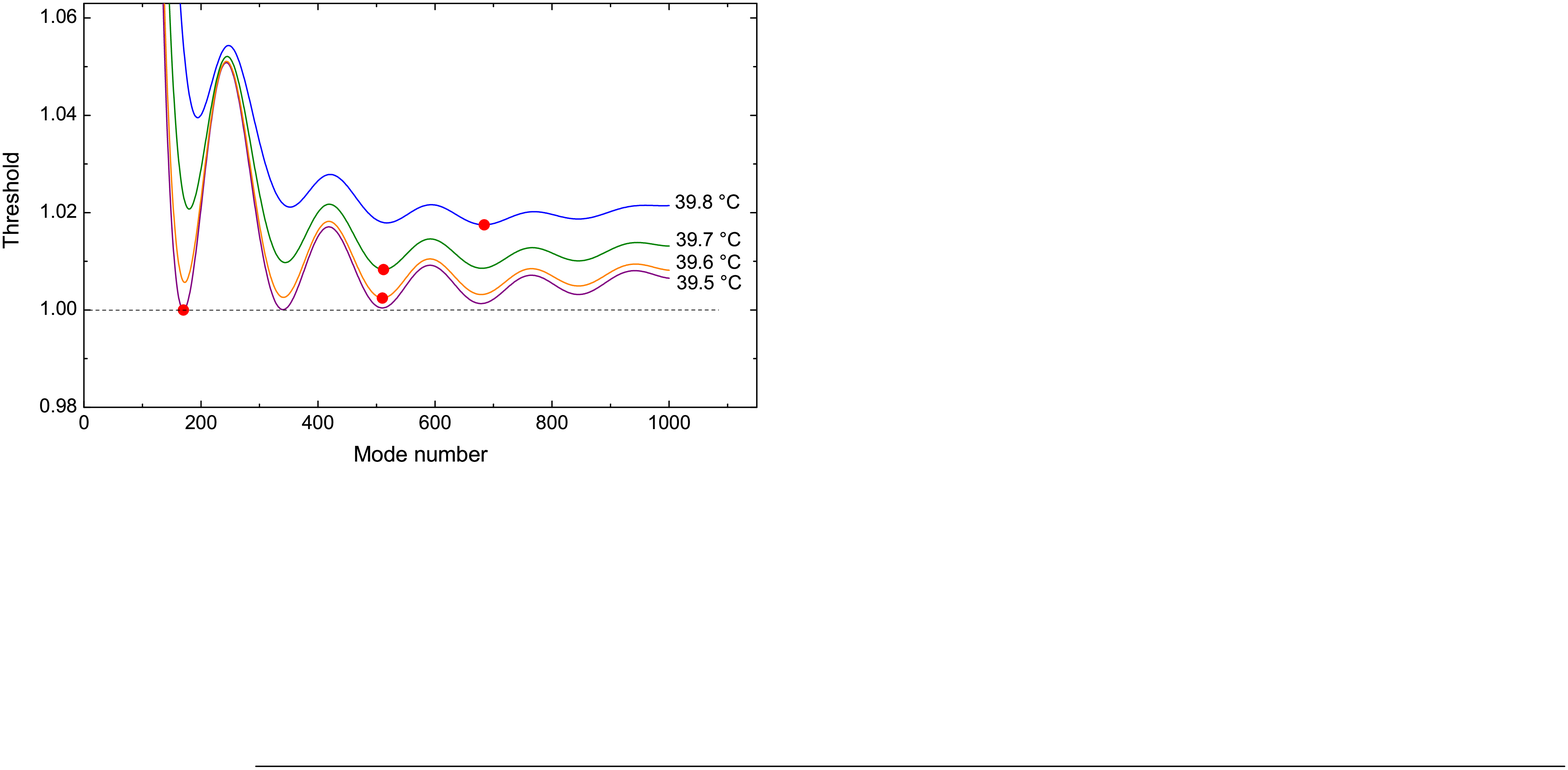}
\caption{Linear stability analysis. Power threshold as a function of the number mode $\mu$ for different temperatures around the SHG phase matching temperature. Red dots indicate the lowest values for each trace.}
\label{fig:stab-analysis1}
\end{center}
\end{figure}

The zero solution for the parametric fields becomes unstable, thus side modes start to oscillate, when the real part of at least one matrix  eigenvalue 
\begin{align}
\lambda_\pm &= -  \gamma - g_0 (\eta_{\mu00\mu}+ \eta_{\bar{\mu}00\bar{\mu}}^*) |a_0|^2
\pm 
\Big\{ g_0^2 |\eta_{\mu\bar{\mu}00}|^2 \, |a_0|^4
\nonumber \\
&- \left[ \Delta_{0}\! - \!D_2 \mu^2 -\! i g_0 (\eta_{\mu00\mu}-\! \eta_{\bar{\mu}00\bar{\mu}}^*) |a_0|^2\right]^2 \Big\}^{1/2}
\label{eq:eigenvalues}
\end{align}
is positive.
For each $\mu$-pair, a power threshold $|a_\text{0,thr}^\mu|^2$ for the intracavity fundamental field  can be found by solving the equation $Re[\lambda]=0$ in the variable $|a_0|^2$. The $\mu$-pair with the lowest threshold is the first to reach a stable oscillation for the cascaded OPO.
A similar expression has been recently derived in a time-domain description of $\chi^{(2)}$ comb dynamics~\cite{Leo:xxxx}. 

An extended discussion of stability analysis is out of the scope of the present review and is in deferred to a future work. 
Figure~(\ref{fig:stab-analysis1}) shows some examples of power threshold as a function of the number mode $\mu$. 
Numerical evaluation has been made using the experimental parameters of Ref.~\cite{Ricciardi:2015bw} and the various traces are calculated for different crystal temperatures around the phase-matching condition for the initial SHG process, with zero pump detuning, $\Delta_0=0$.
When the SHG is phase matched (39.5$^\circ$) the $\mu$-pair with the lowest threshold occurs for $\mu=170$,  even though second and third lowest values, respectively at  $\mu=340$ and 510, differ from the first by less than a part in $10^{4}$.
 We remark that local threshold minima correspond to minima of the losses related to the nonlinear susceptibilities $\eta_{\mu00\mu}$ and $\eta_{\bar{\mu}00\bar{\mu}}$, i.e., SFG processes $\omega_{\mu,\bar{\mu}}+\omega_0$. 
As the crystal is slightly moved from the phase matching condition, the lowest threshold suddenly moves away from the central mode.

From Eq.~(\ref{eq:eigenvalues}) we can derive an approximate expression for the minimum theoretical input power threshold for the onset of the first pair of parametric sidebands. 
If $P$ is the laser power impinging on the nonlinear cavity and ${\cal A}=\pi r^2$ ($r$, cavity mode radius) is the area of a cavity mode, the intensity coupled into the cavity is related to the round-trip coupled field amplitude $F_0$ as
\begin{equation*}
I= \frac{P}{{\cal A}} = \frac{1}{2} \epsilon_0 c \omega_0 \frac{\tau}{2\gamma} |F_0|^2  \, ,
\end{equation*}
where $\epsilon_0$ is the electric constant.
As discussed in Ref.~\cite{Ricciardi:2015bw}, for each possible parametric pair, the threshold value is determined by a trade-off between parametric gain, cavity dispersion, and nonlinear losses. We assume the nonlinear media is phase matched for the SHG of the fundamental laser frequency which, in turn, is perfectly resonant with the central cavity mode $\varpi_0$, i.e., $\Delta_0=0$. Moreover, we neglect the effect of dispersion on cavity modes spacing, thus, each $\mu$-pair is symmetric around the central mode. 
From this symmetry and from the antisymmetry of the imaginary part, Eq.~(\ref{eq:eta_abab2}), the equality $\text{Im}(\eta_{\mu00\mu} - \eta_{\bar{\mu}00\bar{\mu}}^*) = 0$ derives. 
We can further restrain our attention to the theoretical most favorite $\mu$-pairs whose frequencies fall exactly in minima of the most relevant nonlinear losses, given by the SFG processes $\omega_{\mu,\bar{\mu}}+\omega_0$. For these pairs the real part of  $\eta_{\mu00\mu}$ and $\eta_{\bar{\mu}00\bar{\mu}}$ are nil. Therefore, the real part of the eigenvalues reduces to 
$
\lambda_\pm \simeq -  \gamma 
\pm
g_0  |a_0|^2
$,
being $|\eta_{\mu\bar{\mu}00}| \simeq 1$ in the vicinity of the central mode. The intracavity threshold power is then $|a_{0,\text{thr}}|^2 = \gamma/g_0$.
Substituting in Eq.~(\ref{eq:cubic}), and solving for the coupled power $|F_0|^2$, we finally get the input laser power threshold,
\begin{equation*}
P_\text{thr} = \frac{2 \pi \epsilon_0 r^2 c \, n_0^5}{Q^2  d^2}
\end{equation*}
where $Q=\omega_0/\gamma$ is the resonator quality factor.
As an example, for figures typical of state-of-the-art AlGaAs resonators~\cite{Kuo:2014aa,Mariani:2014gw,Pu:2015vt} ($r$ = 0.2~$\mu$m; $n_0 = 3$; $Q = 10^5$; $d$ = 80~V/pm), the singly resonant power threshold is around 4~mW.

\begin{figure}[t!]
\begin{center}
\includegraphics*[bbllx=40bp,bblly=40bp,bburx=700bp,bbury=450bp,width=75mm]{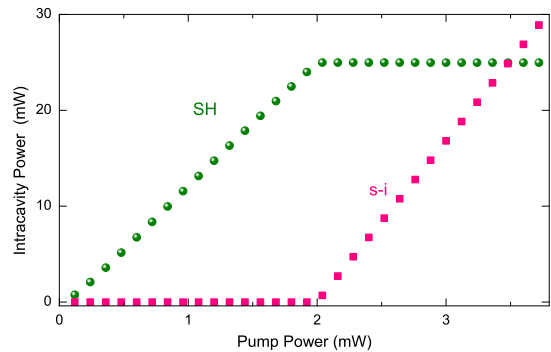}
\caption{Doubly resonant SHG cavity. Numerical simulation of the  intracavity second harmonic and parametric power as a function of the pump power.}
\label{fig:DR}
\end{center}
\end{figure}

We finally observe that a SHG cavity can be simultaneously resonant for both the fundamental and SH frequency. 
For such a doubly resonant SHG cavity, it is reasonable to expect that the pump power threshold of the internally pumped cascaded OPO is lowered with respect to the singly resonant configuration. 
Thus, we numerically solved the coupled mode equations for  SHG and OPO only, with the boundary condition dictated by the presence of the cavity, since the other processes can be neglected in the case of the lowest-threshold oscillating mode.
Fig.~\ref{fig:DR} shows numerical simulation of intracavity  SH and parametric power for cavity parameters as in~\cite{Ricciardi:2015bw}, except that now also the SH field is resonant. We have assumed an outcoupling mirror reflectivity of 0.92, while all the other mirror are perfectly reflecting for the SH. 
The internally pumped OPO threshold is 2~mW, to be compared with the 100~mW threshold measured for the singly resonant SHG cavity.

\section{Final remarks}
Demonstration of direct frequency comb generation in quadratic media is a recent breakthrough revealing a potential for a novel class of highly efficient and versatile frequency comb synthesizers based on $\chi^{(2)}$ nonlinear interactions.
In this review, we have shown the very first steps towards a promising new field of research where most of the work is still to be done.
Exploiting the intrinsically higher efficiency of second-order processes, $\chi^{(2)}$ combs can play a major role in developing small-footprint integrated photonic devices.
Second-order processes have been demonstrated in Si, Si$_3$N$_4$,  AlN, AlGaAs~\cite{Levy:2011wp,Cazzanelli:2012jz,Jung:2014jt,Miller:2014dz,Kuo:2014aa,Mariani:2014gw}. These materials can provide technological platforms for new micro and nanometric devices. 
With respect to Kerr-combs based on $\chi^{(3)}$ materials, $\chi^{(2)}$ combs can be realized all over the transparency range of the nonlinear material without the more severe constraint on the dispersive characteristics of $\chi^{(3)}$-based devices, and  the simultaneous occurrence of octave-distant combs provides a useful metrological link between two spectral regions without the need for a full octave-wide comb. 

However, optimal design of new, more efficient frequency comb synthesizers with lower power threshold and larger bandwidth requires however a thorough analysis of this new phenomena and several questions need to be addressed in order to have a fully developed theoretical framework and reliable predictive models. 
A complementary time-domain approach to $\chi^{(2)}$ comb dynamics  can offer an interesting new perspective, focusing on the time evolution of $\chi^{(2)}$ combs, emphasizing the importance of non-istantaneous character of the nonlinear interaction and the enabling role of temporal walk-off arising from group velocity mismatch~\cite{Leo:xxxx}.
The proposed quantitative spectral dynamical model, here extended to a generic number of interacting modes, offers a deep physical insight into $\chi^{(2)}$ comb dynamics and, interestingly, displays a striking analogy with the description of frequency comb generation in Kerr nonlinear resonators, suggesting a unified theoretical framework for the physics of frequency combs.

\section*{acknowledgement}
We thank Miro Erkintalo, Tobias Hansson, Fran\c{c}ois Leo, and Stefan Wabnitz for stimulating discussions. 
S.~M. acknowledges the Italian “Ministero dell’Istruzione, dell’Universit\`a e della Ricerca” (MIUR) for financial support through FIRB Project n. RBFR13QUVI ``Optomechanical tailoring of squeezed light''.
This work has been partly supported by MIUR Progetto Premiale ``QUANTOM---Quantum Opto-Mechanics''.

\bibliography{../X(2)-comb}

\end{document}